\documentclass{article}
\usepackage{amssymb}
\usepackage{amsmath}
\usepackage{xcolor}

\newcommand{\bea}{\begin{eqnarray}}
\newcommand{\eea}{\end{eqnarray}}
\newcommand{\be}{\begin{equation}}
\newcommand{\ee}{\end{equation}}
\newcommand{\Tr}{\,\textrm{Tr}\,}
\newcommand{\eff}{\,\textrm{eff}\,}
\renewcommand{\top}{\,\textrm{top}\,}
\newcommand{\ground}{\,\textrm{ground}\,}
\newcommand{\ns}{\,\textrm{ns}\,}
\newcommand{\s}{\,\textrm{s}\,}

\newcommand{\rt}[1]{{}}
\newlength{\szovszel}
\newlength{\slashszel}

\begin{document}

\title{Range of topological fluctuations\\ in the three-flavor linear sigma model}
\date{}
\author{Gergely Fej\H os and Andr\'as Patk\'os \\
Institute of Physics and Astronomy, E\"otv\"os University\\
H-1117, P\'azm\'any P\'eter s\'et\'any 1/A, Budapest, Hungary}

%\vfill
\maketitle

\begin{abstract}
The topological charge density two-point function is computed in a three-flavor effective linear meson model, including contributions from topological sectors with arbitrary charge. A corresponding composite field is introduced via a Hubbard-Stratonovich transformation. In the $\eta$–$\eta^\prime$ sector, the model yields very accurate spectroscopy with negligible mixing with the composite field. The resulting topological susceptibility allows for the extraction of a surprisingly large absolute scale for the range of topological fluctuations. When compared with various QCD-based estimates, this indirectly suggests the dominance of higher topologically charged configurations in the ground state.
\end{abstract}

\section{Introduction}
The introduction of dynamically interacting effective degrees of freedom is essential for a physical interpretation of QCD at low energies. A prominent example is the transparent topological characterization of chiral fields, where effective meson models have played a crucial role in elucidating the nature of the $U_A(1)$ anomaly in QCD \cite{thooft86}. A quantitatively satisfactory account of the $\eta$–$\eta^\prime$ mass splitting and the associated topological features was provided by simple effective models that include the topological charge density in the form of an auxiliary field \cite{veneziano79,witten79,divecchia80,rosenzweig80}. The simplest $U_A(1)$-breaking cubic interaction term (commonly known as the ’t Hooft determinant) has become a standard tool for reconstructing the meson mass spectrum in linear effective meson models, dating back to the early 1970s \cite{chan73}. The connection between the anomaly and appropriate decay widths (e.g., that of $\eta \rightarrow 3\pi$) was investigated in \cite{kawarabayashi80,kawarabayashi81}. Further developments in meson spectroscopy were made using improved techniques such as renormalized perturbation theory \cite{lenaghan00,herpay05} and the functional renormalization group \cite{jungnickel95,jungnickel97,mitter14}. The dynamical hadronization strategy developed by Pawlowski and collaborators has significantly advanced the connection between effective quark-meson models and QCD with increasing reliability \cite{pawlowski16,pawlowski20}. Correlators of the topological charge density, as well as the topological susceptibility, can also be accessed through effective models \cite{fukushima01,inagaki11,adhikari24,jiang16}. More recently, attention has turned to the role of higher topologically charged configurations, particularly through the study of higher powers of the ’t Hooft determinant combined with $U_A(3)\times U_V(3)$-invariant operators \cite{fejos24}. Extended meson mass spectra, including vector mesons, have been successfully parametrized in the mean-field approximation by incorporating charge-2 topological operators \cite{parganlija13,kovacs13,kovacs25}. A particularly intriguing line of research has emerged from the proposal that the QCD vacuum may be dominated by configurations with higher topological charge \cite{rennecke24}. Whether the $U_A(1)$ anomaly remains broken above the critical temperature also continues to be of interest \cite{kovacs24,giordano24,tiwari25}.

The first calculations of the topological susceptibility in effective models were performed within a perturbative framework \cite{fukushima01,jiang16}. In these works, the topological charge density correlators of the three-flavor Nambu–Jona-Lasinio model and the linear sigma model were evaluated in the lowest nonvanishing order. While the results were qualitatively satisfactory, the original calculational strategy developed to address the $\eta$–$\eta^\prime$ mass difference \cite{veneziano79,divecchia80}, based on the introduction of an auxiliary field representing the topological charge density, has, to the best of our knowledge, not yet been applied to determine the range of topological charge density fluctuations. The aim of the present study is to fully exploit this simple yet elegant approach, which goes beyond perturbation theory by introducing an effective composite field to represent the topological charge density. By comparing the resulting expression with the Witten–Veneziano relation \cite{veneziano79,witten79,divecchia80}, as well as with recent lattice QCD determinations of the topological susceptibility \cite{borsanyi16,aoki18}, we find compelling evidence pointing to the dominance of higher-charged topological configurations in the $T=0$ vacuum state.

The structure of the paper is as follows. Section 2 summarizes the quantities that characterize the topological features of the effective linear meson model. In this Section, a composite field representing the topological charge density is also introduced via a simplified version of the Hubbard–Stratonovich (HS) transformation. Section 3 presents a detailed analysis of the mass matrix in the extended model, closely following, while slightly extending the original approach of Veneziano. The range of fluctuations of the composite topological field is then derived by integrating out the fields in the $\eta$ sector, resulting in an effective model for the composite field. This result is independently confirmed in Sec. 4 through an exact one-step diagonalization of the full mass matrix. Both approaches consistently indicate a significantly longer range for topological fluctuations compared to the original compositeness scale introduced by the HS transformation. Conclusions are presented in Sec. 5.

\section{Topological susceptibility}

The three-flavor model of pseudoscalar and scalar mesons quite successfully describes the thermodynamics of QCD with couplings fixed to match the $T=0$ spectra. In this Section those quantities and concepts are introduced that are essential for discussing its topological features.

\subsection{ $U_A(1)$ breaking in the linear meson model}

The linear $U_L(3)\times U_R(3)$ symmetric sigma model of the meson fields $M=(s_a+i\pi_a)T^a$ [$\Tr(T^a T^b)=\delta^{ab}/2$] is defined through the Euclidean action
\be
S_M=\int_x\Big[\textrm{Tr}(\partial_{i}M^\dagger\partial_{i}M)+U(I_1,I_2,I_3)-\Tr[H(M+M^\dagger)]\Big], 
\label{action}
\ee
with a potential depending on the group invariants $I_1=\textrm{Tr}(M^\dagger M), I_l=\textrm{Tr}(M^\dagger M-\frac13 I_1)^l,~l=2,3$. This part of the action is also invariant under $U_A(1)$ transformations.
The last term in the right-hand side of (\ref{action}) is responsible for the explicit symmetry breaking, where we assume the form $H=h_0T^0+h_8T^8$. In addition let us consider the most general $U_A(1)$ breaking potential for the three-flavor model,
\be
S_{M,\top}=\sum_{n,i,j,k}A^{(i,j,k)}_n \int_x\left(I_1^iI_2^jI_3^k\right)(\textrm{det}M+\textrm{det}M^\dagger  )^n.
\label{most-general-determinant-function}
\ee
This functional form was investigated recently \cite{rennecke24}, although with a slightly different grouping of the terms. The $U_A(1)$ charge of the determinant terms is the same in both constructs.
The meson matrix transforms under axial $U_A(3)$ as
\be
M\rightarrow A^\dagger M A^\dagger
\ee
with $A=\exp(i\theta_A^aT^a)$, therefore 
\be
\textrm{det}M\rightarrow e^{-3\times i\theta_A^0\times 2/\sqrt{6}}\textrm{det}M
\ee
under $U_A(1)$. The $\Theta$ dependence of the additional term in the action is made explicit as
\be
S_{M,\top}(\Theta)=\sum_{n,i,j,k}A^{(i,j,k)}_n\int d^4x\left(I_1^iI_2^jI_3^k\right)(e^{-i\Theta}\textrm{det}M+e^{i\Theta}\textrm{det}M^\dagger  )^n,\qquad \Theta=\sqrt{6}\theta.
\ee
Promoting $\Theta$ to a spacetime dependent source, the topological charge density is obtained as
\be
Q_M(x)=\frac{\delta S_{M,\top}}{\delta \Theta(x)}\Bigg|_{\Theta=0}=-i\sum_{n,i,j,k}A^{(i,j,k)}_n\left(I_1^iI_2^jI_3^k\right)n\left(\textrm{det}M+\textrm{det}M^\dagger\right)^{n-1}(\textrm{det}M-\textrm{det}M^\dagger).
\ee
One can introduce an effective anomaly coefficient characterizing the ``amplitude'' of $Q_M$:
\bea
Q_M(x)&=&i\tilde{{\cal A}}_{\eff}[M(x)]\big(\textrm{det}M(x)-\textrm{det}M^\dagger (x)\big), \nonumber\\
 \tilde{\cal A}_{\eff}[M(x)]&=&\sum_{n,i,j,k}A^{(i,j,k)}_n\left(I_1^iI_2^jI_3^k\right)n\left(\textrm{det}M+\textrm{det}M^\dagger\right)^{n-1}.
\label{top-charge-density}
\eea
 Note that in the physical vacuum a real condensate of $\langle M \rangle$ is present, therefore $\langle Q_M\rangle =0$, and the topological susceptibility can be defined as
\be
\chi_{M,\top}=\int d^4x \langle Q_M(x) Q_M(0) \rangle,
\label{effective-top-susc}
\ee
where $\langle ... \rangle$ refers to averaging with respect to the action $S_M+S_{M,\top}[\Theta=0]$. In what follows, the average on the right-hand side of (\ref{effective-top-susc}) will be approximated by taking into account only the cubic correlations dictated by the standard ’t Hooft term. Further terms multiplying the fluctuating cubic operators will be evaluated in mean field approximation. Specifically,
\be
\chi_{M,\top} =\tilde{A}^2_{\eff} \int d^4x\Big\langle i(\textrm{det}M(x)-\textrm{det}M^\dagger(x))i(\textrm{det}M(0)-\textrm{det}M^\dagger(0))\Big\rangle,
\ee
where $\tilde{A}_{\eff}$ is the ground state average of $\tilde{\cal A}_{\eff}[M(x)]$ neglecting all correlations. In the ground state at $T=0$, in addition to the spontaneous breaking of $U_V(3)\times U_A(3)$, explicit breaking leads to a diagonal meson matrix $\langle M \rangle \big|_{\textrm{\ground}}=v_{\ns}T^{\ns}+v_{\s}T^{\s}=\textrm{diag}(v_{\ns}/2,v_{\ns}/2,v_{\s}/\sqrt{2})$. Here the index ``ns'' stands for ``non-strange'', ``s'' for ``strange'' \footnote{The transformation between the $0$ and $8$ and ns-s bases is the following:$\quad
\begin{pmatrix}
T^{\ns} \\ T^{\s}
\end{pmatrix}
= \frac{1}{\sqrt3} \begin{pmatrix}
\sqrt2 & 1 \\
1 & -\sqrt2
\end{pmatrix}
\begin{pmatrix}
T^0 \\ T^8
\end{pmatrix}$. }. 
In particular, the coefficient $\tilde{A}_{\eff}$ in the ground state is the following:
\be
\tilde{A}_{\eff}=\sum_{n,i,j,k}A^{(i,j,k)}_n\left(I_1^iI_2^jI_3^k\right)\Big|_{v_{\ns},v_{\s}}n\left(\frac{v_{\ns}^2v_{\s}}{2\sqrt{2}}\right)^{n-1}.
\ee
Introducing $\tilde{Q}_M(x) = i(\textrm{det}M(x)-\textrm{det}M^\dagger(x))$, the topological susceptibility in our approximation becomes
\be
\chi_{M,\top} = \tilde{A}_{\eff}^2 \int d^4x \langle \tilde{Q}_M(x) \tilde{Q}_M(0) \rangle,
\label{effective-top-susc2}
\ee
which coincides with the correlator investigated by Jiang {\it et al.} \cite{jiang16}.

In the next Section we will construct a scalar field corresponding to $Q_M(x)/m_c^3\approx \tilde{A}_{\eff}\tilde{Q}_M(x)/m_c^3 \equiv q_M(x)$, dominating the correlator in the right-hand side of (\ref{effective-top-susc}), where $m_c$ is a characteristic compositeness scale. Under the assumption that the $q_M$ field is propagating freely with mass $m_{\top}$, one finds
\be
\int d^4x \langle Q_M(x)Q_M(0)\rangle=m_c^6\int_x \int_pe^{-ipx}G_{q_M}(p),\qquad G_{q_M}(p)=(p^2+m_{\top}^2)^{-1},
\label{meson-topcharge-correlation}
\ee
where $G_{q_M}$ is the Euclidean propagator of $q_M$. With this the following parametric representation can be considered:
\be
\chi_{M,\top}=m_c^{6}m^{-2}_{\top}.
\label{mesonic-top-susc}
\ee
We see that it is sufficient to produce the static component ($p=0$) of the effective propagator, $G_{q_M}$, of the topological fluctuations. The question to be addressed is the dependence of $m_c$ and $m_{\top}$ on the mass spectra of the model.

\subsection{Extension with a composite field for the topological charge density}

In this Subsection we extend the linear sigma model by introducing a collective field for the topological charge density carried by meson configurations.

Let us introduce the collective field $q_M(x)$ (the rescaled topological charge density) with the usual mass dimension by a trick reminiscent of the HS transformation \cite{jungnickel96,gies02,floerchinger09,jakovac19}. One has the freedom to multiply the ``partition function" of the effective meson model by a mere constant
\be
\int{\cal {D}}q_M\exp\left\{-\int d^4x \Bigg(q_M(x)-\frac{\tilde A_{\eff}\tilde{Q}_M(x)}{m_c^3}\Bigg)\frac{m_c^2}{2}\Bigg(q_M(x)-\frac{\tilde A_{\eff}\tilde{Q}_M(x)}{m_c^3}\Bigg)\right\}.
\label{gauss-constant}
\ee
The scale $m_c$ is yet undetermined, though it can be identified with the quantity that turned out to be identical with $m_{\top}$ in \cite{veneziano79,divecchia80}, since $m_c^2/2$ is the coefficient of the term $q_M^2(x)$. 

The combination $\tilde{Q}_M \equiv i(\det M - \det M^\dagger )$ contains terms that are either linear or cubic in the $\pi^a$ fields. In the broken symmetry phase the linear term certainly dominates, therefore, we omit the cubic ones. The contribution of the latter could be computed perturbatively with the propagators to be determined below. We readily find that the linear piece depends exclusively on $\pi_{\!\ns}\equiv\eta_{\ns}$ and $\pi_{\s}\equiv\eta_{\s}$,
\be
q_M \frac{\tilde A_{\eff}}{m_c}\tilde{Q}_M(x) \equiv iq_M \frac{\tilde A_{\eff}}{m_c}\big(\det M (x)- \det M^\dagger(x) \big) \rightarrow -\frac{\tilde A_{\eff}}{2\sqrt{2}m_c}q_M\left(v_{\ns}^2\eta_s+2v_{\ns}v_{\s}\eta_{\ns}\right).
\label{linear-top-charge-coupling}
\ee
The additional piece that does not depend on $q_M$ (it is quadratic in $\eta_s,\eta_{ns}$) produces an additional contribution to the original (mesonic) mass matrix of this subsector of the sigma model. Altogether one generates the following 3-by-3 matrix for the squared mass in the $q_M$–$\eta_{\ns}$–$\eta_{\s}$ space:
\be
{\cal M}^2 =
\displaystyle
\begin{bmatrix}
m_c^2&\frac{-\tilde A_{\eff}}{\sqrt{2}m_c}v_{\s}v_{\ns}&\frac{-\tilde A_{\eff}}{2\sqrt{2}m_c}v_{\ns}^2\\
\frac{-\tilde A_{\eff}}{\sqrt{2}m_c}v_{\s}v_{\ns}&m^2_{\eta_{\ns,\ns}}+\frac{\tilde A_{\eff}^2}{2m_c^4}v_{\ns}^2v_{\s}^2&m^2_{\eta_{\s,\ns}}+\frac{\tilde A_{\eff}^2}{4m_c^4}v_{\ns}^3v_{\s}\\
\frac{-\tilde A_{\eff}}{2\sqrt{2}m_c}v_{\ns}^2&m^2_{\eta_{\s,\ns}}+\frac{\tilde A_{\eff}^2}{4m_c^4}v_{\ns}^3v_{\s}&m^2_{\eta_{\s,\s}}+\frac{\tilde A_{\eff}^2}{8m_c^4}v_{\ns}^4,
\end{bmatrix}
\label{extended-mass-matrix}
\ee
where $m^2_{\eta_{\ns,\ns}}$, $m^2_{\eta_{\s,\ns}}$, $m^2_{\eta_{\s,\s}}$ are the respective mass squares in the original model. Here one should note that a flavor singlet instanton liquid generates additive contribution to the squared mass of the $\pi_0\equiv \eta_0$ field (see Appendix A of \cite{shuryak95}) the same way as it is  argued in the large $N_{c}$ limit in \cite{veneziano79,divecchia80}. In the present basis such coupling would correspond to the choice $v_{ns}=\sqrt{2}v_{s}$, i.e., in the vacuum $\langle M \rangle \sim {\bf 1}$. This is realized without explicit symmetry breaking terms, meaning that the breaking pattern is $U_L(3)\times U_R(3) \rightarrow U_V(3)$.

\section{Analysis {\it \`a la} Veneziano}
\label{veneziano}
A first approach is to assume that $m_{\top}\ll m_\eta,m_{\eta^\prime}$ and determine the larger masses hierarchically, initially neglecting the impact of the $q_M$ field on the $\eta$ sector. This means that one repeats the analysis of Veneziano \cite{veneziano79} applied to the reduced 2-by-2 mass matrix
\be
{\cal \tilde{M}}^2=
\displaystyle
\begin{bmatrix}
m^2_{\eta_{\ns,\ns}}+\frac{\tilde A_{\eff}^2}{2m_c^4}v_{\ns}^2v_{\s}^2&m^2_{\eta_{\s,\ns}}+\frac{\tilde A_{\eff}^2}{4m_c^4}v_{\ns}^3v_{\s}\\
m^2_{\eta_{\s,\ns}}+\frac{\tilde A_{\eff}^2}{4m_c^4}v_{\ns}^3v_{\s}&m^2_{\eta_{\s,\s}}+\frac{\tilde A_{\eff}^2}{8m_c^4}v_{\ns}^4
\end{bmatrix}
\label{shifted-mass-matrix}
\ee
Below we rely on the commonly accepted treatment of the meson spectra in the framework of the linear sigma model \cite{herpay05,rai20}. For instance, Herpay {\it et al.} \cite{herpay05} used the perturbatively renormalizable terms in the potential of the linear sigma model to derive tree level expressions for all members of the scalar and pseudoscalar meson nonets. This includes only the term linear in $\det M+\det M^\dagger$, usually called the 't Hooft term. When the latter is set to zero, in order to avoid double counting the effect of the anomaly,  the following expressions arise:
\be
m^2_{\eta_{\ns,\ns}}=m_\pi^2,\qquad m^2_{\eta_{\s,\s}}=m_K^2\left(1+\frac{v_{\ns}}{\sqrt{2}v_{\s}}\right)-\frac{v_{\ns}}{\sqrt{2}v_{\s}}m_\pi^2,\qquad m_{\eta_{\s,\ns}}^2=0.
\ee
Since at the physical point $v_{\ns}\approx v_{\s}$ at $T=0$, below we use the unified notation $v_{\ns}\approx v_{\s}\equiv v$.  In addition we use the physical values for the masses  of the pion and kaon (to be denoted by $M_\pi, M_K$), since those are determined from the partially conserved axial-vector current relations and are not sensitive to the $U_A(1)$ anomaly. 
 
\subsection{Diagonalization in the $\eta-\eta^\prime$ sector}
\label{diagonalisation-eta}

After diagonalization, the eigenvalues of the modified mass matrix, denoted by $m_\eta^2$, $m_{\eta^\prime}^2$,  are identified with the physical mass predictions. Denoting the experimental values by $M_\eta^2$, $M_{\eta^\prime}^2$, we set the physical scale by requiring $m_\eta^2+m_{\eta^\prime}^2=M_\eta^2+M_{\eta^\prime}^2$  \cite{veneziano79}. The trace of the mass matrix is invariant, which restricts the parameter of  the Gaussian (\ref{gauss-constant}), $1/m_c^2$, in proportion to $\tilde A_{\eff}$:
\be
\frac{\tilde A_{\eff}^2}{m_c^4}=8\frac{M_\eta^2+M_{\eta^\prime}^2-m_{\eta_{\ns,\ns}}^2-m_{\eta_{\s,\s}}^2}{v_{\ns}^2(v_{\ns}^2+4v_{\s}^2)}\approx \frac85\frac{M_\eta^2+M_{\eta^\prime}^2-M_K^2(1+1/\sqrt{2})-M_\pi^2(1-1/\sqrt{2})}{v^4},
\label{instanton-matching}
\ee
where we also indicated the experimental values for the pion and kaon masses, $m_{\pi}^2 \rightarrow M_{\pi}^2$, $m_{K}^2 \rightarrow M_{K}^2$. It is obvious that our construction makes sense, since the right-hand side of (\ref{instanton-matching}) is positive. Independent information is needed for the separate determination of $\tilde A_{\eff}$ and $m_c$.

We emphasize that the shift in the mass mass matrix comes from topologically nontrivial configurations of the meson fields. At this point we make use of the physical value of the trace in the $\eta$ sector. Using the physical input $M_\eta^2+M_{\eta^\prime}^2\approx 1.2181$ $\textrm{GeV}^2$ with $M_{\pi}\approx 140\textrm{ MeV}$, $M_K \approx 496 \textrm{ MeV}$ the following value emerges from (\ref{instanton-matching}):
\be
\frac{\tilde A^2_{\eff}}{8m_c^4}v^4=\frac{1}{5} (M_\eta^2+M_{\eta^\prime}^2-m_{\eta_{\ns,\ns}}^2-m_{\eta_{\s,\s}}^2)\approx 0.158 \textrm{ GeV}^2.
\ee
Then the mass matrix is fully determined and its diagonalization leads to
\be
m_\eta\approx 538 \textrm{ MeV},\qquad m_{\eta^\prime}\approx 964 \textrm{ MeV}.
\ee
These are to be compared with the PDG data $M_\eta \approx 548\textrm{ MeV}, M_{\eta^\prime}\approx 958\textrm{ MeV}$. (We note that the choice $v_{ns}=\sqrt{2}v_s$ would reproduce the spectra obtained in \cite{veneziano79}.)

\subsection{Naive estimate of the topological susceptibility}
\label{naive}

Before proceeding further, we present an estimate for the topological susceptibility in the crudest approximation, in which the $m_c$ parameter of the HS transformation is identified with the topological mass, $m_c=m_{\top}$. For the second term in the right-hand side of (\ref{mesonic-top-susc}), one has
\be
\displaystyle
\chi_{M,\top}=\frac{m_c^6}{m_{\top}^2}=m_c^4.
\label{top-susc}
\ee
One might question the compatibility of this result with the Witten-Veneziano relation:
\be
m_c^4=\frac{1}{6}v^2(M_\eta^2+M_{\eta^\prime}^2-2M_K^2).
\ee
Making use of (\ref{instanton-matching}) one finds the following expression for $\tilde A_{\eff}$:
\be
\tilde A_{\eff}^2=\frac{4}{15v^2}(M_\eta^2+M_{\eta^\prime}^2-2M_K^2)\big(M_\eta^2+M_{\eta^\prime}^2-M_K^2(1+1/\sqrt{2})-M_\pi^2(1-1/\sqrt{2})\big).
\label{A-eff-expr}
\ee
A recent computation using the functional renormalization group with regard to the scalar and pseudoscalar meson spectra with the choice of [cf. (\ref{most-general-determinant-function})]
\be
S_{M,\top}=\int_x \Big[A_1(I_1)(\textrm{det}M+\textrm{det}M^\dagger  )+A_2(I_1)(\textrm{det}M+\textrm{det}M^\dagger  )^2 \Big]
\ee
leads to $|\tilde A_{\eff}|\approx A_1(I_1\big|_\textrm{ground})+2A_2(I_1\big|_\textrm{ground})\big(\textrm{det}\langle M\rangle|_{\textrm{ground}}+\textrm{det}\langle M^\dagger \rangle|_{\textrm{ground}}\big) \approx 4.9\textrm{ GeV}$ \cite{fejos24}. When substituting the physical data, choosing $v\approx 93 \textrm{ MeV}$ in (\ref{A-eff-expr}), one finds $|\tilde A_{\eff}|\approx 4.21\textrm{ GeV}$, which appears as satisfactory evidence for the consistency of this simplified treatment. However, the analysis of the next Subsection will challenge the validity of the relation $m_{\top}=m_c$.

\subsection{Reduced effective model for $q_M$}

A standalone effective model for the topological fluctuations can be derived by integrating over $\eta$ and $\eta^\prime$. The diagonalization of the mass matrix presented in Sec. \ref{diagonalisation-eta} can be realized via a rotation by angle $\phi$ in the two-dimensional $(\eta_s,\eta_{ns})$ plane, which leads to the mass terms
\be
\frac{1}{2}\left(m_{\eta^\prime}^2\eta^{\prime2}+m_\eta^2\eta^2\right)
\label{diagonal-mass-eta-sector}
\ee
in the Lagrangian. The eigenmodes $\eta,\eta^\prime$ are given by
\be
\eta^\prime=\eta_{\ns}\cos\phi+\eta_{\s}\sin\phi,\qquad \eta=\eta_{\s}\cos\phi-\eta_{\ns}\sin\phi,
\ee
where
\be
\tan(2\phi)=\frac{2{\cal \tilde{M}}^2_{{\s,\ns}}}{{\cal \tilde{M}}_{{\ns,\ns}}^2-{\cal \tilde{M}}^2_{{\s,\s}}}\approx 7.13 ~~\rightarrow~~\phi\approx 41.0^{\circ}.
\ee
If one relates the rotation to the $\eta_0$–$\eta_8$ basis, one finds for the rotation angle about $\delta\approx -14^{\circ}$, which agrees quite well with the observed mixing angle $-11.5^{\circ}$, as well as the result of \cite{rennecke17}.

One also expresses the term linear in $q_M$ in (\ref{linear-top-charge-coupling}), in terms of the eigenmodes of the mass matrix,
\bea
&\displaystyle
- q_M(d_{\eta^\prime}\eta^\prime+d_\eta\eta),
\label{eta-linear-term}
\eea
where
\bea
d_{\eta^\prime}=\frac{\tilde A_{\eff}v^2}{2\sqrt{2}m_c}(\sin\phi+2\cos\phi),\qquad 
d_\eta=\frac{\tilde A_{\eff}v^2}{2\sqrt{2}m_c}(\cos\phi-2\sin\phi).
\eea
For the integration over $\eta$ and $\eta^\prime$ one has to complete the sum of (\ref{diagonal-mass-eta-sector}) and (\ref{eta-linear-term}) into a full square, which induces a negative contribution to the mass term of $q_M$. The resulting estimate for the range of the topological fluctuations modifies according to the following expression:
\be
\mu_c^2=m_c^2\left(1-\frac{d^2_\eta}{m_\eta^2}-\frac{d_{\eta^\prime}^2}{m_{\eta^\prime}^2}\right).
\ee
By substituting the numerical values obtained above, the shifted mass coefficient, $m_c^2\rightarrow \mu_c^2$, of $q_M^2$ is obtained as,
\be
\mu_c^2 \approx 0.029m_c^2.
\label{hierarchy-mu-m}
\ee
This is radically smaller than the naive estimate and leads to 
\be
\chi_{M,\top}\approx 34.21 m_c^4.
\ee
One might object that the hierarchical calculation, which neglects the influence of the composite field on the spectra in the $\eta_{\ns}$–$\eta_{\s}$ sector might not be legitimate. Therefore, we turn back to the exact diagonalization of the extended 3-by-3 mass matrix, given in (\ref{extended-mass-matrix}).

\section{Exact solution of the extended model}

Let us go back to (\ref{extended-mass-matrix}) and fully diagonalize it, including the $q_M$–$\eta$ elements. A more transparent representation is possible (when choosing $v_{\s}\approx v_{\ns}\equiv v$) using the combination 
\be
X=\frac{\tilde A_{\eff}}{2\sqrt2 m^2_c}v^2.
\ee
Then the ${\cal M}^2$ mass matrix takes the following form:
 \be
 {\cal M}^2=
\begin{bmatrix}
m_c^2&-2Xm_c&-Xm_c\\
-2Xm_c&m_{\eta_{\ns,\ns}}^2+4X^2&2X^2\\
-Xm_c&2X^2&m_{\eta_{\s,\s}}^2+X^2
\label{extended-mass-matrix-B}
\end{bmatrix} .
\ee
From this representation it is clear that the condition for neglecting the $q_M$–$\eta$ mixing is simply $X \gg m_c$, which should be checked {\it a posteriori}.
Denoting the exact eigenvalues of ${\cal M}^2$ as $\mu_c^2,\mu_\eta^2,\mu_{\eta^\prime}^2$, we can write the characteristic polynomial as
\be
(\mu_c^2-\lambda)(\mu_\eta^2-\lambda)(\mu_{\eta^\prime}^2-\lambda).
\label{chpol}
\ee
One can determine the eigenvalues using the three equations obtained by comparing (\ref{chpol}) with the original form of the characteristic polynomial,
\bea
&\displaystyle
\mu_c^2+\mu_\eta^2+\mu_{\eta^\prime}^2=m_c^2+m_{\eta_{\ns,\ns}}^2+m_{\eta_{\s,\s}}^2+5X^2\nonumber\\
&\displaystyle
\mu_c^2(\mu_\eta^2+\mu_{\eta^\prime}^2)+\mu_\eta^2\mu_{\eta^\prime}^2=m^2_{\eta_{\s,\s}}m^2_{\eta_{\ns,\ns}}+m_c^2(m_{\eta_{\s,\s}}^2+m_{\eta_{\ns,\ns}}^2)+X^2(4m^2_{\eta_{\s,\s}}+m_{\eta_{\ns,\ns}}^2)\nonumber\\
&\displaystyle
\mu_c^2\mu_\eta^2\mu_{\eta^\prime}^2=m_c^2m_{\eta_{\s,\s}}^2m_{\eta_{\ns,\ns}}^2.
\label{cubic-eigenvalue}
\eea
Following \cite{veneziano79}, as was also done in the previous section, we partially set the physical scale by requiring the calculated trace in the $\eta$ sector to match the observed value.
\be
\mu_\eta^2+\mu_{\eta^\prime}^2=- \mu_c^2+m_c^2+m_{\eta_{\ns,\ns}}^2+m_{\eta_{\s,\s}}^2+5X^2=M_\eta^{2}+M_{\eta^\prime}^{2}\equiv S.
\label{trace-condition}
\ee
Then, from the second equation one can express $\mu_\eta^2\mu_{\eta^\prime}^2$ as a linear combination of $\mu_c^2$ and $m_c^2$ (after eliminating $X^2$ with help of the previous equation),
\bea
\mu_\eta^2\mu_{\eta^\prime}^2&\!\!\!=\!\!\!&\mu_c^2\left(-S+\frac{1}{5}(m_{\eta_{\ns,\ns}}^2+4m_{\eta_{\s,\s}}^2)\right)+\frac{1}{5}m_c^2(4m_{\eta_{\ns,\ns}}^2+m_{\eta_{\s,\s}}^2) \nonumber\\
&&-\frac{1}{5}(m_{\eta_{\ns,\ns}}^4+4m_{\eta_{\s,\s}}^4)+\frac{1}{5}S(4m_{\eta_{\s,\s}}^2+m_{\eta_{\ns,\ns}}^2)\equiv F.
\label{defF}
\eea
Substituting (\ref{defF}) into the left-hand side of the third equation of (\ref{cubic-eigenvalue}), one arrives at a quadratic equation for $\mu_c^2$ with a non-negative discriminant,
\bea
&&\mu_c^2\Big[\mu_c^2\left(-S+\frac{1}{5}(m_{\eta_{\ns,\ns}}^2+4m_{\eta_{\s,\s}}^2)\right)+\frac{1}{5}m_c^2(4m_{\eta_{\ns,\ns}}^2+m_{\eta_{\s,\s}}^2)\nonumber\\
&&\hspace{2.2cm}-\frac{1}{5}(m_{\eta_{\ns,\ns}}^4+4m_{\eta_{\s,\s}}^4)+\frac{1}{5}S(4m_{\eta_{\s,\s}}^2+m_{\eta_{\ns,\ns}}^2)\Big]=m_c^2m_{\eta_{\s,\s}}^2m_{\eta_{\ns,\ns}}^2.
\label{quadratic-mu2}
\eea
Only the root, which vanishes for $m_c^2\rightarrow 0$ can be considered physical and it will determine the range of topological fluctuations (i.e., $\mu_c$) as a function of $m_c^2$. Using it in (\ref{trace-condition}) one also finds $X^2$,  that is $\tilde A_{\eff}$ as function of $m_c^2$. Eventually, with the parametric knowledge of $\mu_c^2(m_c^2)$, one can also reconstruct the quadratic equation, which determines $m_\eta^2$ and $m_{\eta^\prime}^2$ separately,
\be
m_\eta^4-Sm^2_\eta+F=0.
\ee
It is natural to expect that the smaller of the two solutions of (\ref{quadratic-mu2}) allows one to test the validity of the result of the hierarchical integration derived in the previous section. One finds it by choosing the negative sign in the quadratic formula of the equation. To do it directly, we assume that to leading order $\mu_c^2 \sim m_c^2$, and both are much smaller than the mass scale of the $\eta$ sector, $S=1.2181\textrm{ GeV}^2$. That is,  in (\ref{quadratic-mu2}) one can neglect the terms containing $\mu_c^4$ and $\mu_c^2m_c^2$. The resulting linear relation already has only one root, which satisfies the expected proportionality to $m_c^2$,
\be
\mu_c^2\left(-\frac{1}{5}(m_{\eta_{\ns,\ns}}^4+4m_{\eta_{\s,\s}}^4)+\frac{1}{5}S(4m_{\eta_{\s,\s}}^2+m_{\eta_{\ns,\ns}}^2)\right)=m_c^2m_{\eta_{\s,\s}}^2m_{\eta_{\ns,\ns}}^2.
\ee
After substituting the numerical values of $S, m_{\eta_{\s,\s}}^2$ and $m_{\eta_{\ns,\ns}}^2$, it gives
\be
\mu_c^2=0.0296m_c^2. 
\ee
This is quite close to (\ref{hierarchy-mu-m}) and confirms that the spatial  range of topological fluctuations is much larger than the HS scale. It is radically different from the classic approach outlined in Sec. \ref{naive}. In this limiting case one finds
\be
\chi_{M,\top}=\frac{m_c^6}{\mu_c^2}\approx 33.78 m_c^4
\ee
for the topological susceptibility. One chooses the value of $m_c^2$ to arrange reliable estimation for $\chi_{M,\top}$. For the validity of the Witten-Veneziano formula one has to choose $m_c^2=5.56\times 10^{-3}\textrm{ GeV}^2$. Matching results from lattice simulations of full QCD \cite{borsanyi16,aoki18,petreczky16} or its $N_f=2$ version \cite{dimopoulos19} would lead to even lower $m_c^2$ values.
This scale is several order of magnitude smaller than $S$, confirming the validity of the perturbative treatment of the solution with respect to $m_c^2$.

The roots of the characteristic equation of the $\eta$ sector  
to first order in $m_c^2$ are
\be
m_\eta^2=0.2865\textrm{ GeV}^2-0.1486m_c^2,\qquad m_{\eta^\prime}^2=0.9316\textrm{ GeV}^2+0.1486m_c^2.
\ee
The corrections to the result of the $m_c^2=0$ case are certainly fairly small. Therefore, the mixing of $q_M$ with $\eta$ and $\eta^\prime$ is negligible, though definitely nonzero. 

With the Witten-Veneziano input one finds just slight modifications in $m_\eta^2,m_{\eta^\prime}^2$ relative to those found in Sec. \ref{naive},
\be
m_\eta=538\textrm{MeV},\qquad m_{\eta^\prime}=964\textrm{MeV}.
\ee
These values change very little when a lower estimate is used for $\chi_{M,\top}^{1/4}$.

\section{Discussion and conclusion}

In this paper, a complete topological characterization of the effective three-flavor meson model has been presented in the broken phase of chiral symmetry. A composite field, $q_M(x)$, representing the $U_A(1)$ breaking topological charge density, entered the discussion in two essential aspects. First, it provided a shift in the mass matrix of the $\eta$–$\eta'$ sector, which resulted in very accurate mass spectra. Second, it has been demonstrated that the integration over the $\eta,\eta^\prime$  fields has a significant impact on the two-point function of $q_M(x)$, specifically, the spatial range of topological fluctuations increases to more than five times the compositeness scale  [see (\ref{hierarchy-mu-m})]. 

This investigation can be further completed by perturbatively computing the two-loop contributions to the $q_M$ correlator arising from the coupling of three pseudoscalar fields to the effective composite degree of freedom, which is omitted in the present investigation. At zero temperature this contribution was shown to be negligible in Ref. \cite{jiang16}. At finite temperature, however, its weight is expected to increase due to the decreasing vacuum condensates.

In order to keep the topological susceptibility at a fixed, ``physical'' value, the chosen compositeness mass scale should be reduced, which results in $\tilde A_{\eff}$ being one-fifth the strength of the 't Hooft coupling, $A_1$, determined from the effective model description of the mesonic spectrum. This apparent tension could be relaxed by assuming large destructive contributions from the higher charged $U_A(1)$ breaking operators to $\tilde A_{\eff}$. This pattern was proposed recently in Ref. \cite{rennecke24}.

Finally, we note that off-shell (fluctuation) effects may also play an important role and could be computed, for example, using the functional renormalization group technique. In this approach, one establishes a flow equation for the scale-dependent quantum effective action including also the anomalous term (\ref{most-general-determinant-function}), which must be integrated from the ultraviolet (UV) down to the infrared (IR) scales, thereby incorporating all fluctuations. When evaluating the effective action at the IR scale, one can repeat the procedure outlined in this paper, i.e., separating the $U(3)\times U(3)$ invariant part and introducing $q_M$ as an auxiliary variable to enable the calculation of topological characteristics. Such a procedure can shed light on the importance of fluctuation effects in relation to the anomaly, but we leave this for future work.

\section*{Acknowledgments}
G.F. was supported by the Hungarian National Research, Development, and Innovation Fund under Project No. FK142594. A.P. was supported by the Hungarian National Research, Development, and Innovation Fund under Project No.  K143460.

\end{document}